\documentclass[aps,prb,twocolumn,showpacs,floatfix,pdf]{revtex4}

\usepackage{amsmath}    % need for subequations
\usepackage{graphicx}   % need for figures
\usepackage{verbatim}   % useful for program listings
\usepackage{color}      % use if color is used in text
\usepackage{subfigure}  % use for side-by-side figures
\usepackage{hyperref}   % use for hypertext links, including those to external documents and URLs
\usepackage{amssymb}
\usepackage{amsfonts}
\usepackage{tabularx}
\usepackage{dcolumn}% Align table columns on decimal point
\usepackage{bm}% bold math
\usepackage{floatflt}
\usepackage{setspace}
\usepackage{epstopdf}

\usepackage{subfigure}
\usepackage{natbib}

\begin{document}
\title{Investigation of on-site inter-orbital single electron hoppings in general multi-orbital systems}
%\title{Importance of on-site inter-orbital single electron hoppings in multi-orbital systems}
%\title{Presence of the on-site inter-orbital single hopping in multi-orbital systems}
%\title{The on-site inter-orbital hopping effects in multi-orbital systems}

\author{Qingguo Feng}
\email[]{qingguo.feng@physics.uu.se}
\affiliation{Department of Physics and Astronomy, Uppsala University, Box 516, S-75120 Uppsala, Sweden}
\author{P.\ M.\ Oppeneer}
\affiliation{Department of Physics and Astronomy, Uppsala University, Box 516, S-75120 Uppsala, Sweden}

\date{\today}

\begin{abstract}
A general multi-orbital Hubbard model, which includes on-site inter-orbital electron hoppings, is introduced and studied.
It is shown that the on-site inter-orbital single electron hopping is one of the most basic interactions. Two electron spin-flip and pair-hoppings are shown to be correlation effects of higher order
than the on-site inter-orbital single hopping. It is shown how the double and higher hopping interactions can be well-defined for arbitrary systems. The two-orbital Hubbard model is studied numerically to demonstrate the influence of the single electron hopping effect, leading to a change of the shape of the bands and a shrinking of the difference between the two bands. Inclusion of the on-site inter-orbital hopping suppresses the so-called orbital-selective Mott transition.
\end{abstract}

% insert suggested PACS numbers in braces on next line \pacs{}
%71.10.-w 	Theories and models of many-electron systems
%71.10.Fd 	Lattice fermion models (Hubbard model, etc.)
%71.15.-m 	Methods of electronic structure calculations
%71.27.+a 	Strongly correlated electron systems; heavy fermions
%71.30.+h 	Metal-insulator transitions and other electronic transitions
%\keywords{Impurity solver, Green's function, Dynamical mean field theory, Metal-insulator transition, Hubbard model}

\maketitle

Strongly correlated electron systems have drawn great interest, as they exhibit many exotic phenomena, such as metal-insulator transitions, unusual forms of magnetism, superconductivity, and heavy-fermion behavior.\cite{Grewe,ImadaRMP1998} These systems are therefore investigated theoretically in both model and {\it ab initio} calculations. One of the most essential tasks is to model the system appropriately, i.e., such that a realistic physical picture can be attained.
In condensed matter theory the Hubbard model, which was originally proposed in the early sixties,\cite{Hubbard63,Gutzwiller63,Kanamori63} is one of the simplest, yet also the most important and most frequently studied lattice model to investigate strongly correlated electron systems. It sets up a competition between an inter-site quantum mechanical hopping term and an on-site Coulomb interaction term. As a consequence the model can describe various non-trivial phenomena. Due to its simplicity and because the model has captured the essence
of strongly correlated electron systems, the Hubbard model has been widely used.\cite{RevHubbard1,citeHubbard,citeHubbard2,citeHubbard3}\\
\indent
In a realistic situation an atom in a correlated system will usually have several partial filled orbitals and should therefore be described with a multi-orbital (MO) Hubbard model. With the inclusion of orbital degrees of freedom, inter-orbital interactions have to be included for such a system.
The inter-site hoppings should now be a sum of the inter-site intra-orbital hoppings and the inter-site inter-orbital hoppings, where the two kinds of hopping are defined according to the orbital indices of the start and the end orbitals for hopping electron are identical or not, respectively.
Moreover, besides the remain competition of the inter-site hopping and on-site Coulomb interactions, one new kind of on-site interactions will join into this competition, i.e., the on-site inter-orbital hoppings. Acting as the inter-site hoppings, the on-site inter-orbital hoppings
are also associated to the kinetic energy of electrons. All the inter-site hoppings, on-site inter-orbital hoppings and on-site Coulomb interactions (which also divides as on-site intra-orbital Coulomb interactions and on-site inter-orbital Coulomb interactions) constitute a competition for various interactions existing in a realistic multi-orbital system.
%In previously studies, the usually studied multi-orbital Hubbard model \cite{Werner09,Liebsch,Medici,HTSC1} in a general form is  %\cite{Kuzemsky98}
%\begin{eqnarray}
%{\cal H}&=&-\sum_{ijlm\sigma,i\neq j}t_{ijlm}f^{\dag}_{il\sigma}f_{jm\sigma}
%+\sum_{il}U_{ll}\hat{n}_{il\uparrow}\hat{n}_{il\downarrow}\nonumber\\
%&&+\sum_{ilm\sigma\sigma',l<m}U_{lm\sigma\sigma'}\hat{n}_{il\sigma}\hat{n}_{im\sigma'}\nonumber\\
%&&-J\sum_{ilm,l\neq m,\sigma\neq\sigma'}f^{\dag}_{im\sigma'}f^{\dag}_{il\sigma}f_{il\sigma'}f_{im\sigma}\nonumber\\
%&&+J\sum_{ilm,l\neq m,\sigma\neq\sigma'}f^{\dag}_{im\sigma'}f^{\dag}_{im\sigma}f_{il\sigma'}f_{il\sigma},\label{Hubbardmodel}
%\end{eqnarray}
%where $i, j$ are site indices, $l, m$ are orbital indices, $\sigma, \sigma'$ are spin indices. The $f^{\dag}_{il\sigma}, f_{il\sigma}$ are correspondingly the generate and destroy operator for $\sigma$ spin in $l$-th orbital on site $i$. The first term is the inter-site hopping term, where $t_{ijlm\sigma}$ is the hopping amplitude for spin $\sigma$ hopping from $m$-th orbital on site $j$ to $l$-th orbital on site $i$. The second term is the intra-orbital Coulomb interaction term, and the third term is the inter-orbital Coulomb interaction term, where $U_{lm\uparrow\downarrow}=U_{lm\uparrow\uparrow}+J$ and $J$ is the Hund's coupling constant. Moreover, the last two terms are the so-called spin-flip exchange term and the pair-hopping term, which are considered that these two terms are important for the studies of the high-temperature superconductivity.\cite{HTSC1}
%To do this in an appropriate, realistic manner
Therefore, for an arbitrary multi-orbital system, we propose a generalized MO-Hubbard model as follows,
%%Hamiltonian (to distinguish from the usually studied general form of the Hubbard model, we call this the multi-orbital Hubbard-Feng model)
%shown as
\begin{eqnarray}
{\cal H}&=&-\sum_{ijlm\sigma,i\neq j}t_{ijlm}f^{\dag}_{il\sigma}f_{jm\sigma}
+\sum_{il}U_{ll}\hat{n}_{il\uparrow}\hat{n}_{il\downarrow}\nonumber\\
&&+\sum_{ilm\sigma\sigma',l<m}U_{lm\sigma\sigma'}\hat{n}_{il\sigma}\hat{n}_{im\sigma'}\nonumber\\
&&+\sum_{ilm\sigma,l<m}\big(I^{\ast}_{ilm\sigma}f^{\dag}_{im\sigma}f_{il\sigma}^{~}+I_{ilm\sigma}f^{\dag}_{il\sigma}f^{~}_{im\sigma}\big),\label{Hubbard-Feng}
\end{eqnarray}
where $i, j$ are site indices, $l, m$ are orbital indices, and $\sigma, \sigma'$ are spin indices. The $f^{\dag}_{il\sigma}, f_{il\sigma}$ are the creation and annihilation operator, respectively, for spin $\sigma$ in $l$-th orbital on site $i$. The first term is the inter-site hopping term, where $t_{ijlm\sigma}$ is the hopping amplitude for spin $\sigma$ hopping from $m$-th orbital on site $j$ to $l$-th orbital on site $i$. The second term is the intra-orbital Coulomb interaction and the third term is the inter-orbital Coulomb interaction, where $U_{lm\uparrow\downarrow}$=$U_{lm\uparrow\uparrow}+J$ and $J$ is the Hund's coupling constant.
The last two terms in the Hamiltonian are the on-site inter-orbital single electron direct hopping terms,  where the $I^{\ast}_{ilm\sigma}$ and  $I_{ilm\sigma}$ are the hopping amplitudes, as $t_{ijlm}$ in the first term.
Note that this part is written in a form of one term and its conjugate, so that one can easily observe that the system is in a statistical equilibrium state. For the half filled case, $I_{ilm\uparrow}=I_{ilm\downarrow}=I^{\ast}_{ilm\uparrow}=I^{\ast}_{ilm\downarrow}$, the model automatically obeys spin rotational invariance.
Here the on-site inter-orbital single electron hopping is to be considered as one basic interaction along with the inter-site hopping and the on-site Coulomb interaction.
One may argue that with specially chosen basis the on-site inter-orbital hopping will disappear. However, we are here discussing a general multi-orbital system with arbitrary basis. It is convenient to investigate such a system because in some experiments the natural chosen basis in a material may not be orthogonal. Thus studying a system theoretically with a same basis set as in the experimental material can easily gives the results comparative to the physical observables obtained in experiments. Moreover, we find that the inclusion of on-site inter-orbital single hoppings has introduced some new and interesting theory.\\
\indent
%
%Here we show that the on-site inter-orbital single electron hoppings are to be considered as one of the most basic on-site inter-orbital interactions along with the on-site Coulomb interactions. They describe a realistic inter-orbital process $f^{\dag}_{m\sigma}(t')f_{l\sigma}(t)$ that one electron is annihilated in the $l$-th orbital at time $t$ and one electron with identical spin is created in the $m$-th orbital at time $t'$ where $l$$\neq$$m$.
Recently we have numerically studied \cite{FO11a,FO11b} the first three terms in Eq.\ \eqref{Hubbard-Feng} within the dynamical mean field theory (DMFT).\cite{DMFT1,DMFT2}
%It was observed that the inclusion of the inter-site inter-orbital hoppings and the on-site inter-orbital fluctuations suppresses the so-called orbital selective Mott transition (OSMT) \cite{OSMT,OSMT1} in a two-orbital system having different band widths for the two orbitals.
In this work we formulate theoretically and evaluate numerically the influence of the on-site inter-orbital single electron hopping.
Using the  DMFT the proposed generalized Hubbard model can be mapped to a generalized single impurity Anderson model (SIAM) along with a self-consistency condition. The Hamiltonian of this generalized SIAM is
\begin{eqnarray}
{\cal H}_{imp}&=&\sum_{kl\sigma}\varepsilon_{kl\sigma}c^{\dag}_{l k\sigma}c_{l k\sigma}^{~}+\sum_{l\sigma}\varepsilon_{fl\sigma} f^{\dag}_{l\sigma}f_{l\sigma}+\sum_{l}U_{ll}\hat{n}_{l\uparrow}\hat{n}_{l\downarrow}\nonumber\\
&+&\sum_{lm\sigma\sigma',l<m}U_{lm\sigma\sigma'}\hat{n}_{l\sigma}\hat{n}_{m\sigma'}\nonumber\\
&+&~~\sum_{l k\sigma}~~~~\big(V^{\ast}_{l k\sigma}c^{\dag}_{l k\sigma}f_{l\sigma}+V_{l k\sigma}f^{\dag}_{l\sigma}c_{l k\sigma}\big)\nonumber\\
&+&\sum_{lm\sigma,l<m}~\big(I^{\ast}_{lm\sigma}f^{\dag}_{m\sigma}f_{l\sigma}+I_{lm\sigma}f^{\dag}_{l\sigma}f_{m\sigma}\big),
\label{SIAMmodel}
\end{eqnarray}
where the first term is the energy of the conduction electrons (bath), $c^{\dag}_{l k\sigma}$ and $c_{l k\sigma}$ are correspondingly the creation and annihilation operators of conduction electrons. The second term is the energy of the localized electrons. The third (fourth) term represents the intra-orbital (inter-orbital) Coulomb interactions. The fifth summation is the hybridization term that gives the interaction between the bath and the localized electrons. The sixth summation represents just the on-site inter-orbital single electron direct hoppings on the impurity site, where for convenience we have dropped the site indices in the symbols.
These inter-orbital single electron hoppings are on-site interactions which do not change in the mapping. We can hence study these terms in Eq.\ \eqref{Hubbard-Feng} by equivalently studying the mapped SIAM.
\\
\indent
We use the equation of motion (EOM) method to solve the DMFT impurity problem. In this method we compute the equations of motion % with the SIAM Hamiltonian Eq.\ \eqref{SIAMmodel}
according to the equation
\begin{eqnarray}
\omega\ll \! A;B \! \gg=\langle[A,B]_+\rangle+\ll \! [A,{\cal H}_{imp}];B \! \gg,
\label{eq:4}
\end{eqnarray}
where we have used the Fourier transform of the double time temperature-dependent retarded Green's function (GF) $\ll \! A(t');B(t) \! \gg$, i.e., $\ll \! A;B \! \gg$ is defined in $\omega$ space.\cite{Zubarev}  $[,]_+$ means the anti-commutator and $[,]$ the commutator. The first term on the right hand side (RHS) labels the interaction associated to $\ll \! A;B \! \gg$, the second term describes the involvement of the higher-order interactions, where the higher-order GFs will appear. Calculating the EOM of these higher-order GFs,  GFs of even higher-order will appear in the newly derived EOMs. Repeating this procedure, more higher and even higher order GFs will appear. Each GF is associated with an {\it order} and one physical interaction. The {\it order} of the GF approximately labels the weight of its associated interaction. Approximately, the  GF of lower order associated with an interaction will give a larger contribution than higher-order GFs associated with the interaction.\\
\indent
This statement can be demonstrated considering the following procedure. The $1^{\rm st}$ order EOM is% calculated as,
%\begin{widetext}
\begin{eqnarray}
&&(\omega+\mu-\varepsilon_{fm\sigma})\ll \! f_{m\sigma};f^{\dag}_{m\sigma} \! \gg=\nonumber\\
&&\qquad\qquad 1+U_{mm}\ll \! \hat{n}_{m\sigma'}f_{m\sigma};f^{\dag}_{m\sigma} \! \gg\nonumber\\
&&\qquad\qquad +\sum_{l,l\neq m}\big(U_{lm\sigma\sigma}\ll \! \hat{n}_{l\sigma}f_{m\sigma};f^{\dag}_{m\sigma} \! \gg\nonumber\\
&&\qquad\qquad\qquad~~~ +U_{lm\sigma'\sigma}\ll \! \hat{n}_{l\sigma'}f_{m\sigma};f^{\dag}_{m\sigma} \! \gg\big)\nonumber\\
&&\qquad\qquad +~\sum_{k}~V_{mk\sigma}\ll \! c_{mk\sigma};f^{\dag}_{m\sigma} \! \gg\nonumber\\
&&\qquad\qquad -\sum_{l,l\neq m}I_{lm\sigma}\ll \! f_{l\sigma};f^{\dag}_{m\sigma} \! \gg,
\label{eq:6}
\end{eqnarray}
%\begin{eqnarray}
%(\omega+\mu-\varepsilon_{km\sigma})\ll c_{mk\sigma};f^{\dag}_{m\sigma}\gg&=&{\cal V}^{\ast}_{mmk\sigma}\ll f_{m\sigma};f^{\dag}_{m\sigma}\gg+\sum_{l}{\cal V}^{\ast}_{mlk\sigma}\ll f_{l\sigma};f^{\dag}_{m\sigma}\gg_{l\neq m}.\label{eq:7}
%\end{eqnarray}
where the last term is generated by the on-site inter-orbital single electron hopping terms, and $\mu$ is the chemical potential.
Next, we calculate the second-order EOMs of those newly appeared higher-order GFs on the RHS of Eq.\ \eqref{eq:6}, e.g., the EOM of the GF $\ll \! f_{l\sigma};f^{\dag}_{m\sigma} \! \gg$ is
\begin{eqnarray}
&&(\omega+\mu-\varepsilon_{fl\sigma})\ll \! f_{l\sigma};f^{\dag}_{m\sigma} \! \gg
=\nonumber\\
&&\qquad\qquad\langle [f_{l\sigma},f^{\dag}_{m\sigma}]_{\dag} \rangle+U_{ll}\ll \! \hat{n}_{l\sigma'}f_{l\sigma};f^{\dag}_{m\sigma} \! \gg\nonumber\\
&&\qquad\qquad+\sum_{l',l'\neq l}\big(U_{l'l\sigma\sigma}\ll \! \hat{n}_{l'\sigma}f_{l\sigma};f^{\dag}_{m\sigma} \! \gg\nonumber\\
&&\qquad\qquad\qquad~~~+U_{l'l\sigma'\sigma}\ll \! \hat{n}_{l'\sigma'}f_{l\sigma};f^{\dag}_{m\sigma}\! \gg\big)\nonumber\\
&&\qquad\qquad+~\sum_{k}~V_{lk\sigma}\ll \! c_{lk\sigma};f^{\dag}_{m\sigma} \! \gg\nonumber\\
&&\qquad\qquad-\sum_{l',l'\neq l}I_{l'l\sigma}\ll \! f_{l'\sigma};f^{\dag}_{m\sigma} \! \gg ,
\label{eq:9}
\end{eqnarray}
%\end{widetext}
where we can observe from the derivative procedure that the last term actually reflects a physically consequent double hopping interaction, $f^{\dag}_{l\sigma}(t_2)f_{l'\sigma}(t_2)f^{\dag}_{m\sigma}(t_1)f_{l\sigma}(t_1)$.\\
\indent
%Moreover, in the second line of Eq.\ \eqref{eq:9}, when $l'=m$, the first GF should be $\ll\hat{n}_{m\sigma'}f_{l\sigma};f^{\dag}_{m\sigma}\gg$, while the second GF is $\ll\hat{n}_{m\sigma}f_{l\sigma};f^{\dag}_{m\sigma}\gg$.
%
Now calculating further the third-order EOMs of the GFs appearing on the RHS of Eq.~\eqref{eq:9}, and taking $\ll \! \hat{n}_{m\sigma'}f_{l\sigma};f^{\dag}_{m\sigma} \! \gg$ as an illustration we obtain\small
%\begin{widetext}
\begin{eqnarray}
&&(\omega+\mu-\varepsilon_{fl\sigma})\ll \! \hat{n}_{m\sigma'}f_{l\sigma};f^{\dag}_{m\sigma} \! \gg
=\nonumber\\
&&\qquad\qquad\langle [\hat{n}_{m\sigma'}f_{l\sigma},f^{\dag}_{m\sigma}]_{+}\rangle+U_{lm\sigma'\sigma}\ll \! \hat{n}_{m\sigma'}f_{l\sigma};f^{\dag}_{m\sigma} \! \gg\nonumber\\
&&\qquad\qquad+U_{ll}\ll \! \hat{n}_{l\sigma'}\hat{n}_{m\sigma'}f_{l\sigma};f^{\dag}_{m\sigma} \! \gg\nonumber\\
&&\qquad\qquad+U_{lm\sigma\sigma}\ll \! \hat{n}_{m\sigma}\hat{n}_{m\sigma'}f_{l\sigma};f^{\dag}_{m\sigma}\! \gg\nonumber\\
&&\qquad\qquad+\sum_{l',l'\neq(l,m)}(U_{l'l\sigma'\sigma}\ll \! \hat{n}_{l'\sigma'}\hat{n}_{m\sigma'}f_{l\sigma};f^{\dag}_{m\sigma}\! \gg\nonumber\\
&&\qquad\qquad+U_{l'l\sigma\sigma}\ll \! \hat{n}_{l'\sigma}\hat{n}_{m\sigma'}f_{l\sigma};f^{\dag}_{m\sigma} \! \gg)\nonumber\\
&&\qquad\qquad+~~~\sum_k~~~\big(-V^{\ast}_{mk\sigma'}\ll \! c^{\dag}_{mk\sigma'}f_{m\sigma'}f_{l\sigma};f^{\dag}_{m\sigma} \! \gg\nonumber\\
&&\qquad\qquad~~~~~~~~~~~~~~~+ V_{lk\sigma}\ll \! \hat{n}_{m\sigma'}c_{lk\sigma};f^{\dag}_{m\sigma} \! \gg\nonumber\\
&&\qquad\qquad~~~~~~~~~~~~~~~+ V_{mk\sigma'}\ll \! f^{\dag}_{m\sigma'}c_{mk\sigma'}f_{l\sigma};f^{\dag}_{m\sigma} \!  \gg\big)\nonumber\\
&&\qquad\qquad - I^{*}_{lm}\ll \! f^{\dag}_{l\sigma'}f_{m\sigma'}f_{l\sigma};f^{\dag}_{m\sigma} \! \gg\nonumber\\
&&\qquad\qquad - \sum_{l',l'\neq (l,m)}I^{*}_{l'm}\ll \! f^{\dag}_{l'\sigma'}f_{m\sigma'}f_{l\sigma};f^{\dag}_{m\sigma} \! \gg\nonumber\\
&&\qquad\qquad+I_{lm}\ll \! f^{\dag}_{m\sigma'}f_{l\sigma'}f_{l\sigma};f^{\dag}_{m\sigma} \! \gg\nonumber
%\\
\end{eqnarray}
\begin{eqnarray}
&&\qquad\qquad+\sum_{l',l'\neq (l,m)}I_{l'm}\ll \! f^{\dag}_{m\sigma'}f_{l'\sigma'}f_{l\sigma};f^{\dag}_{m\sigma} \! \gg\nonumber\\
&&\qquad\qquad+I_{lm}\ll \! \hat{n}_{m\sigma'}f_{m\sigma};f^{\dag}_{m\sigma} \! \gg\nonumber\\
&&\qquad\qquad+\sum_{l',l'\neq (l,m)}I_{l'l}\ll \! \hat{n}_{m\sigma'}f_{l'\sigma};f^{\dag}_{m\sigma} \! \gg,\label{eq:10}
\end{eqnarray}\normalsize
%\end{widetext}
where $l'$$\neq$$(l,m)$ means $l'$$\neq$$ l$ and $l'$$\neq$$m$. The last six terms on the RHS are generated by the on-site inter-orbital single electron hoppings.\\
\indent
Next, noting that
\begin{eqnarray}
&&[f_{m\sigma},f^{\dag}_{m\sigma'}f^{\dag}_{m\sigma}f_{l\sigma'}f_{l\sigma}]= - f^{\dag}_{m\sigma'}f_{l\sigma'}f_{l\sigma},\\ % pair-hopping
&&[f_{m\sigma},f^{\dag}_{l\sigma'}f^{\dag}_{m\sigma}f_{m\sigma'}f_{l\sigma}]= - f^{\dag}_{l\sigma'}f_{m\sigma'}f_{l\sigma},   % spin-flip exchange
\end{eqnarray}
we recognize that $\ll\! f^{\dag}_{m\sigma'}f_{l\sigma'}f_{l\sigma};f^{\dag}_{m\sigma}\! \gg$ in Eq.\ \eqref{eq:10} is actually associated with the pair-hopping term,\cite{KogaPRB05,Werner09} while $\ll\! f^{\dag}_{l\sigma'}f_{m\sigma'}f_{l\sigma};f^{\dag}_{m\sigma}\! \gg$ corresponds to the so-called spin-flip exchange term.\cite{KogaPRB05,Werner09}
One can note that both the pair-hopping and spin-flip exchange terms can be reproduced by on-site inter-orbital single hopping terms. The pair hopping is thus a special type of double hopping in which both the two electrons in the $l$-th orbital hop to the $m$-th orbital simultaneously. Similarly, the spin-flip exchange term is a certain kind of double hopping, in which one particle with spin $\sigma$ hops from the $l$-th orbital to the $m$-th orbital as well as one particle with spin $\sigma'$ hops from $m$-th orbital to $l$-th orbital at the same time, where $\sigma$$\neq$$\sigma'$. %In previous theory these two terms were introduced by the rotational invariance and defined only for two-orbital systems. Here
Thus we have given them a more general definition for arbitrary multi-orbital systems.
Notably, in a multi-orbital system the possible double hoppings are not only the pair-hopping and the spin-flip exchange term. The other four terms in the last six terms on the RHS of Eq.\ \eqref{eq:10} also correspond to certain forms of on-site inter-orbital double hoppings at an equal time limit.\\
\indent
If we continue the procedure to calculate higher-order EOMs, more and higher-order GFs will appear which are  associated with higher-order interactions. In view of the above, the on-site inter-orbital single hoppings $f^{\dag}_{m'\sigma}(t'_1)f_{m\sigma}(t_1)$ are the most basic physical interactions in Hubbard-like strongly correlated systems. These will furnish the higher-order double-time double-hopping interactions, $f^{\dag}_{l'\sigma}(t'_2)f^{\dag}_{m'\sigma'}(t'_1)f_{l\sigma}(t_2)f_{m\sigma'}(t_1)$, that is,
one single hopping occurs at time $t_1$ and another hopping occurs at time $t_2$.
In the same manner
higher-order hoppings, as three-time three-hopping interactions and four-time four-hopping interactions {\it etc.} will appear.
These multi-hopping interactions are embedded in our formulation and appear in higher-order EOMs.
\begin{figure}[tb]
\includegraphics[width=6.5cm,angle=0]{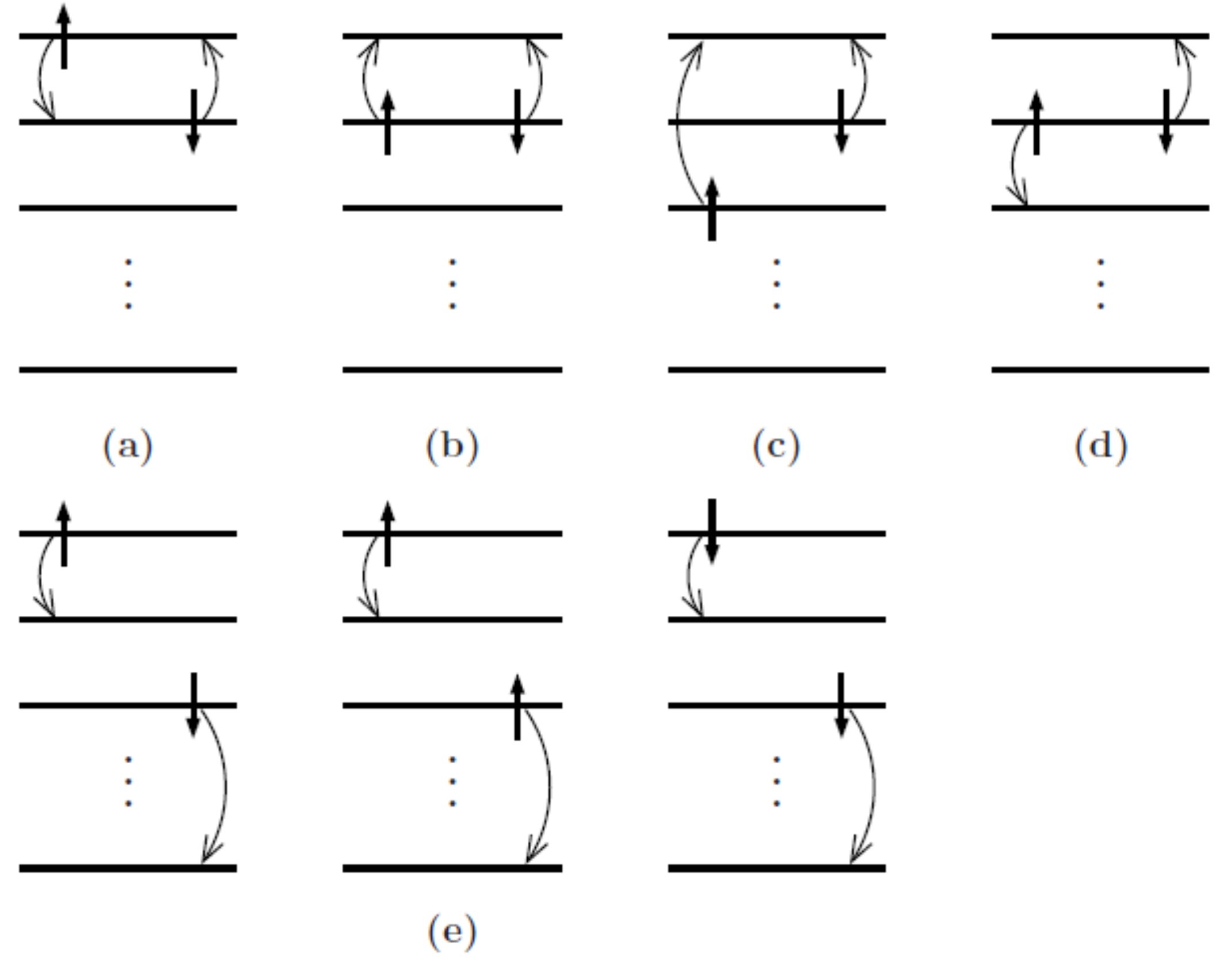}%
\caption{Illustration of possible simultaneous double hoppings in a multi-orbital system.
The straight arrows label the spin up or down of an electron and the bend arrow gives the hopping direction. (a) the spin-flip exchange term;  (b) the pair-hopping term; (c) two electrons in different orbitals with different spins hop to the same orbital; (d) two electrons in one orbital with different spins hop to different orbitals; (e) two electrons in different orbitals hop to different orbitals.
\label{fig1}}
\end{figure}

Next, we investigate in more detail the double-time double hopping at equal times, i.e., the two hoppings occur simultaneously. The general form for the double hoppings is
%\begin{eqnarray}
$f^{\dag}_{l'\sigma}f^{\dag}_{m'\sigma'}f_{l\sigma}f_{m\sigma'}$, %\label{eq:doublehopping}
%\end{eqnarray}
where $l'$$\neq$$ l$ and $m'$$\neq $$m$. This equation gives all the double hoppings that possibly exist in a multi-orbital system. When $l'$=$m'$, $l$=$m$, $\sigma'$$\neq$$\sigma$, it is the pair-hopping between orbitals.\cite{KogaPRB05,Werner09} When $l'$=$m$, $l$=$m'$, $\sigma'$$\neq$$\sigma$, it is the spin-flip exchange term. The whole set of on-site double hoppings include several kinds of interactions. If separated according to initial and final orbital of the hopping, we can schematically represent them as shown in Fig.\ \ref{fig1}. Writing %out (\ref{eq:doublehopping}),
%{\bf there is repetition here ..}
%
%\begin{widetext}
%\begin{eqnarray}
%f^{\dag}_{l'\sigma}f^{\dag}_{m'\sigma'}f_{l\sigma}f_{m\sigma'}
%&=&\sum_{ml\sigma\sigma',m\neq l,\sigma\neq\sigma'}f^{\dag}_{m\sigma}f^{\dag}_{l\sigma'}f_{l\sigma}f_{m\sigma'}
%+\sum_{ml\sigma\sigma',m\neq l,\sigma\neq\sigma'}f^{\dag}_{l\sigma}f^{\dag}_{l\sigma'}f_{m\sigma}f_{m\sigma'}\nonumber\\
%&+&~\sum_{mll',l'\neq l,\sigma'\neq\sigma}(f^{\dag}_{m\sigma}f^{\dag}_{m\sigma'}f_{l'\sigma}f_{l\sigma'}+f^{\dag}_{l'\sigma}f^{\dag}_{l\sigma'}f_{m\sigma}f_{m\sigma'})\nonumber\\
%&+&\sum_{(ll'mm')~all~different}f^{\dag}_{l'\sigma}f^{\dag}_{m'\sigma'}f_{l\sigma}f_{m\sigma'},\label{10}
%\end{eqnarray}
%\end{widetext}
\begin{eqnarray}
&&f^{\dag}_{l'\sigma}f^{\dag}_{m'\sigma'}f_{l\sigma}f_{m\sigma'}=\sum_{ml\sigma\sigma',m\neq l,\sigma\neq\sigma'}f^{\dag}_{m\sigma}f^{\dag}_{l\sigma'}f_{l\sigma}f_{m\sigma'}\nonumber\\
&&~~~+\sum_{ml\sigma\sigma',m\neq l,\sigma\neq\sigma'}f^{\dag}_{l\sigma}f^{\dag}_{l\sigma'}f_{m\sigma}f_{m\sigma'}\nonumber\\
&&~~~+~\sum_{mll',l'\neq l,\sigma'\neq\sigma}(f^{\dag}_{m\sigma}f^{\dag}_{m\sigma'}f_{l'\sigma}f_{l\sigma'}+f^{\dag}_{l'\sigma}f^{\dag}_{l\sigma'}f_{m\sigma}f_{m\sigma'})\nonumber\\
&&~~~+\sum_{(ll'mm')\, {\it all \, different}}f^{\dag}_{l'\sigma}f^{\dag}_{m'\sigma'}f_{l\sigma}f_{m\sigma'},\label{10}
\end{eqnarray}
where the first term is just the spin-flip term and the second term is the pair-hopping term. Note that the fifth term can split into three sub-terms according to the two spins present in different orbitals. And, Fig.~\ref{fig1} only gives an illustration of the double hoppings, the terms should sum over all orbitals. If there are less orbitals, the number of terms will be reduced accordingly. For three-hopping,  four-hopping, and even higher-order multi-hopping interactions, a schematic representation similar to Fig.~\ref{fig1} can be made.\\
\indent
From the above discussions, it can be recognized that %the spin-flip and pair-hopping terms are only two of the possible equal-time inter-orbital double-hopping interactions.
% and the equal-time inter-orbital double-hopping interactions are among a more general theory of the inter-orbital double-time double-hopping interactions.
%Moreover, these
the usually studied spin-flip and pair-hopping terms are higher-order consequences of the on-site inter-orbital single electron hoppings and Coulomb interactions. As the on-site inter-orbital single hopping is of lower order it will have a larger weight (or probability) than the spin-flip and pair-hopping terms.
%and, consequently, it can not be neglected when studying these double hoppings.
We note that, from the Eqs.\ \eqref{eq:6} and \eqref{eq:9}, the off-diagonal GFs $\ll \! f_{l\sigma};f^{\dag}_{m\sigma} \! \gg$ relate to the on-site inter-orbital single hopping. If one neglects the on-site inter-orbital single hopping but includes the pair-hopping and spin-flip terms in a model Hamiltonian, the obtained off-diagonal GFs will be inexact.
Furthermore, in any self-consistent theory where the number of orbitals is larger than two, more inter-orbital double hopping terms have to be included,
% besides the spin-flip term and the pair-hopping term,
as shown by Eq.\ \eqref{10}. Such  terms can be well treated in our theory, because they will be automatically generated from the on-site inter-orbital single hoppings.\\
\indent
The hopping amplitudes $I_{lm\sigma}$ have the physical meaning of the hopping amplitudes. They are to be determined for each studied system, but not fixed.
%Once known they define the prefactors for the spin-flip term ($-I_{mm' \sigma}U I_{mm' \sigma'}$) and the pair-hopping term ($I_{mm' \sigma}U I_{mm' \sigma'}$). These prefactors have then the same unit as the Coulomb interaction, one being negative and one positive, which is consistent with previous theory. \cite{KogaPRB05, Werner09,threeband1}
%However, they are not simply fixed as assumed in previously studied MO-Hubbard model,\cite{KogaPRB05, Werner09,threeband1} but will vary with $I_{mm' \sigma}$ and the studied system.
For example, if in a certain system the $\sigma$ spin channel is fully occupied for one orbital all the incoming single hoppings of spin $\sigma$ into this orbital will be suppressed. In a paramagnetic system one can consider, for simplicity,  that each orbital has the same incoming and out-going inter-orbital single hoppings. %Once the hopping amplitudes have been specified, the generalized Hubbard model can be studied numerically.
Importantly, all studies of higher-order hopping terms must be made on top of having the on-site inter-orbital single hoppings included. %The spin-flip and pair hopping terms and possible higher-order terms cannot be introduced without these single hopping terms.

As an example we have numerically studied the two-orbital Hubbard model in paramagnetic case to show the influence of the on-site inter-orbital hopping. This interaction was explored in Ref.\ \onlinecite{KogaPRB05}, but not in a self-consistent manner. Theoretically, in the situation where the neighboring sites are identical to the impurity site, the influence of the on-site inter-orbital hopping is similar to that of the inter-site inter-orbital hopping, as the orbitals on the neighboring sites are  identical to the orbitals on the impurity site. In order to cleanly distinguish the contribution of the on-site inter-orbital hopping from that of the inter-site inter-orbital hopping, we set the inter-site inter-orbital hopping and on-site inter-orbital fluctuations to zero. We used the MO-EOM impurity solver of Ref.\ \onlinecite{FO11a} in combination with genetic algorithm techniques.\cite{FZJ09,phdthesis-feng}
%{\red check! some explanation of the hopping parameters is needed.}
%
\begin{figure}[t!]
\vspace{0mm}
\includegraphics[width=5.4cm,angle=-90]{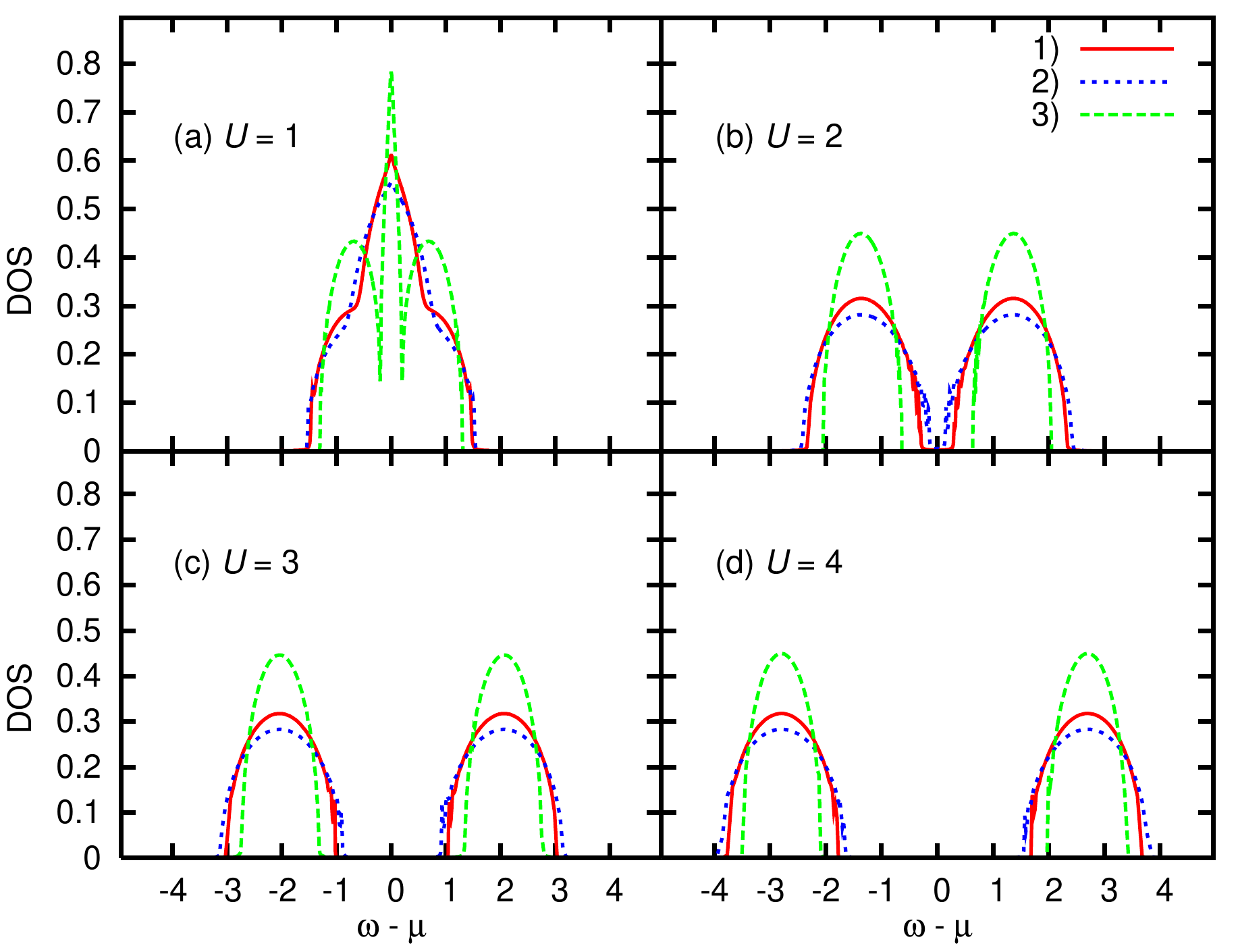}%
\caption{Quasiparticle DOS for the paramagnetic half-filled two-orbital Hubbard model on the Bethe lattice, computed for the case of identical band widths. The parameters used are the half bandwidth $D_2$=$D_1$=1, temperature $T$=0.01, and the Coulomb $U$ as specified in the panels. Line 1) is obtained
for the inter-site intra-orbital hopping parameters $t_2$=$t_1$=$D_1/2$ and no on-site inter-orbital hopping. Line 2) uses the same parameters as 1) but the on-site inter-orbital hopping $t_{12}$=$t_1/2$. Line 3) uses the parameter $t^{tot}_1$=$t_1+t_{12}$=$D_1/2$ and $t_1$=$t_{12}$. \label{fig2}}
\end{figure}
\begin{figure}[t!]
\vspace{2mm}
\includegraphics[width=6.4cm,angle=-90]{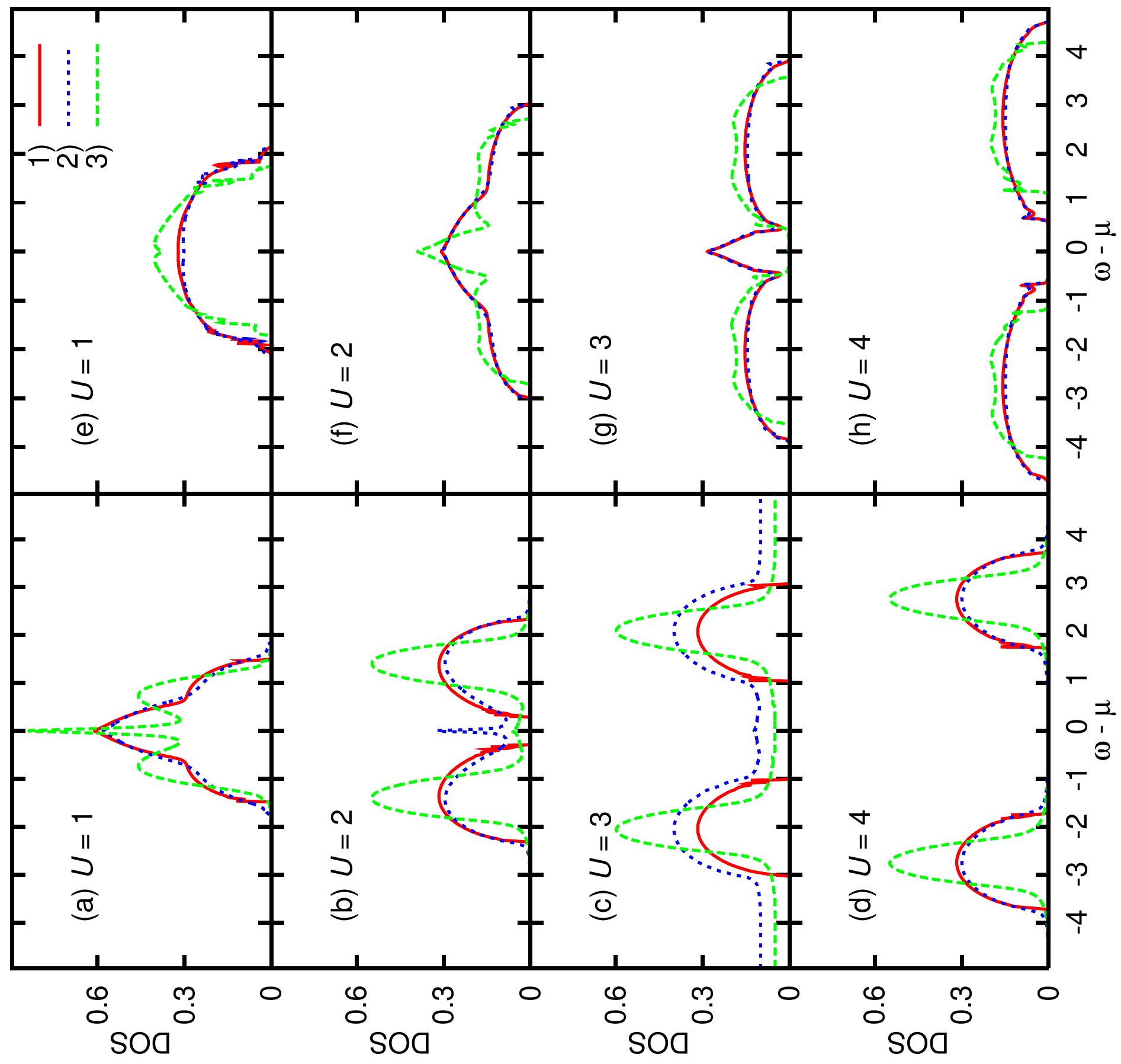}%
\caption{As Fig.\ \ref{fig2}, but for the case of different band widths.
%Quasiparticle densities of states for the paramagnetic half-filled two-orbital Hubbard model on the Bethe lattice.
The left panels depict the DOS for the narrow orbital and the right panels that for the wide orbital. The parameters used are $D_2$=$2D_1$=2 and $T$=0.01. The lines 1) - 3) are obtained as explained in Fig.\ \ref{fig2}, only $t_2$ is modified correspondingly with $D_2$.\cite{note1} The narrow orbital's DOS for lines 2) and 3) with $U$=3 in (c) have been shifted vertically for visibility by 0.1 and 0.05, respectively.
%Line 1) is obtained using the MO-EOM impurity solver in Ref.\ \onlinecite{FO11a} by neglecting the inter-site inter-orbital hoppings and taking the parameters $t_2=2t_1=D_1$. Line 2) is similar to 1) but with the inclusion of an additional on-site inter-orbital hopping $T_{12}=t_1/2$. Line 3 is similar to 2) but with the parameter $t^{tot}_{1}=t_1+t_{12}$ equal to the $t_1$ in 1) and $t^{tot}_{2}=t_2+t_{12}$ equal to the $t_2$ in 1).
\label{fig3}}
\end{figure}

The computed quasiparticle densities of states (DOS) are shown in Figs.\ \ref{fig2} and \ref{fig3} for the cases that the two orbitals have identical or different band widths, respectively. Line 1) is calculated without on-site inter-orbital hopping, line 2) with the same inter-site intra-orbital hopping and a nonzero on-site inter-orbital hopping, and line 3) with both inter-site intra-orbital and on-site inter-orbital hopping but the total hopping amplitude equals that for line 1).\cite{note1}
Line 1) in Fig.\ \ref{fig2} illustrates that there is a metal-insulator transition controlled by $U$. Line 2) shows that
% in the absence of on-site inter-orbital hopping, there is an orbital selective Mott transition.
adding the on-site inter-orbital hopping interaction the Hubbard bands are somewhat broadened. This happens because  the on-site inter-orbital hopping effectively increases the total hopping amplitude so that the electrons gain some itineracy. In Fig.\ \ref{fig3} line 1) shows that in the absence of on-site inter-orbital hoppings there is an orbital selective Mott transition (OSMT).\cite{OSMT} However, from both lines 2) and 3) one observes that the narrow and wide orbitals simultaneously change from metallic states to insulating states along with the increase of $U$. Therefore, the OSMT shown with line 1) is suppressed with the inclusion of the on-site inter-orbital hopping.

To summarize, we introduced a general MO-Hubbard model that can readily be employed to study a broad range of multi-orbital systems.
%, complete, and self-consistent multi-orbital Hubbard model has been introduced to describe the multi-orbital systems.
%It can be employed to study a more broad range of systems. For example, a two-orbital magnetic system having only spin up in each orbital can not be described with the spin-flip term and the pair-hopping term with fixed prefactor, but can be well treated in our introduced multi-orbital Hubbard model.
%
We have shown analytically and numerically the influence of introducing the on-site inter-orbital single hoppings. When these exist in a correlated electron system they will greatly change its properties.
Higher-order effects, such as spin-flip exchange and double hoppings, have to be studied on top of the on-site inter-orbital single hoppings, which, as outlined, can be done in a well-defined way for arbitrary correlated systems.
%As all the higher order correlation effects have to be studied on top of the situation to involving the single hoppings,
The developed theory is expected to be beneficial for studies of unsolved correlation-related phenomena and to trigger inspiring theoretical studies and discoveries.\\
\indent
%
%Finally, we would like to mention the meaning of introducing the on-site inter-orbital direct single hopping that it gives a correct picture of the multi-orbital systems. In the multi-orbital Hubbard model, the pair-hopping terms need an extreme situation that the out-going orbital must be doubly occupied and the destination orbital must be empty. While in the multi-orbital Hubbard-Feng model, the inter-orbital single hopping can be existed for a large range of systems, where whether an inter-orbital single hopping is suppressed or encouraged depends on the states of the system, so that it may explain more phenomena that can not be explained with the multi-orbital Hubbard model, e.g., the single electron excitation. From the theory, we are confident that the Hubbard-Feng model will be beneficial to the studies of superconductivity, and will trigger more interesting theoretical studies and discoveries.
%
We thank the Swedish Research Council (VR) and SKB for financial support. Computer time from the Swedish National Infrastructure for Computing (SNIC) is acknowledged.


\begin{thebibliography}{99}
\bibitem{note1}
Specifically, the parameters used in Fig.\ \ref{fig2} are: line 1) $t_1$=$t_2$=$D_1/2$, $t_{12}$=$t_{21}$=0; line 2)  $t_1$=$t_2$=$D_1/2$, $t_{12}$=$t_{21}$= $D_1/4$; line 3)  $t_1$=$t_2$=$D_1/4$, $t_{12}$=$t_{21}$=$D_1/4$.
The parameters used in Fig.\ \ref{fig3} are: line 1) $t_1$=$D_1/2$, $t_2$=$D_2/2$, $t_{12}$= $t_{21}$=0; line 2)  $t_1$=$D_1/2$, $t_2$=$D_2/2$, $t_{12}$=$t_{21}$=$D_1/4$; line 3)  $t_1$=$D_1/4$, $t_2=D_2/2 - D_1/4$, $t_{12}$=$t_{21}$=$D_1/4$.
\end{thebibliography}
\end{document}